\documentstyle[12pt]{article}
\newlength{\extraspace}
\newlength{\extraspaces}
\newcommand{\be}{\begin{equation}
\addtolength{\abovedisplayskip}{\extraspaces}
\addtolength{\belowdisplayskip}{\extraspaces}
\addtolength{\abovedisplayshortskip}{\extraspace}
\addtolength{\belowdisplayshortskip}{\extraspace}}
\newcommand{\ee}{\end{equation}}

\newcommand{\ba}{\begin{eqnarray}
\addtolength{\abovedisplayskip}{\extraspaces}
\addtolength{\belowdisplayskip}{\extraspaces}
\addtolength{\abovedisplayshortskip}{\extraspace}
\addtolength{\belowdisplayshortskip}{\extraspace}}
\newcommand{\ea}{\end{eqnarray}}

\newcommand{\nonu}{\nonumber \\[.5mm]}
\newcommand{\A}{&\!\!\!}

\newcommand{\newsection}[1]{
\vspace{5mm} \pagebreak[3] \addtocounter{section}{1}
\setcounter{subsection}{0} \setcounter{footnote}{0}
\begin{center}
{\large {\bf \thesection. #1}}
\end{center}
\nopagebreak
\medskip
\nopagebreak \hspace{3mm}}

\begin{document}

\pagenumbering{arabic}

\begin{center}
{{\bf Reissner Nordstr$\ddot{o}$m solutions and Energy in
teleparallel theory}}
\end{center}
\centerline{ Gamal G.L. Nashed}

\bigskip

\centerline{{\it Mathematics Department, Faculty of Science, Ain
Shams University, Cairo, Egypt }}

\bigskip
 \centerline{ e-mail:nasshed@asunet.shams.edu.eg}

\hspace{2cm}
\\
\\
\\
\\
\\
\\
\\
\\

We give  three different spherically symmetric spacetimes for the
coupled gravitational and electromagnetic fields with charged
source in the tetrad  theory of gravitation. One of these,
contains an arbitrary function and generates the others. These
spacetimes give the Reissner Nordstr$\ddot{o}$m metric black hole.
We then, calculated the energy associated with these spacetimes
using the superpotential method. We find that unless the
time-space components of the tetrad field go to zero faster than
${1/\sqrt{r}}$ at infinity, one gets different results for the
energy.

\newpage
\begin{center}
\newsection{\bf Introduction}
\end{center}

Energy-momentum, angular momentum and electric charge play central
roles in modern physics. The conservation of the first two is
related to the homogeneity and isotropy of spacetime respectively
while charge conservation is related to the invariance of the
action integral under internal U(1) transformations. Local
quantities such as energy-momentum, angular momentum and charge
densities are well defined if gravitational fields are not present
in the system. However, in general relativity theory a
well-behaved energy-momentum and angular momentum densities have
not yet been defined, although total energy-momentum and total
angular momentum can be defined for an asymptotically flat
spacetime surrounding an isolated finite system. The equality of
the gravitational mass and the inertial mass holds within the
framework of general relativity \cite{MTW, SNK,KST}. However, such
equality is not satisfied for the Schwarzschild metric when it is
expressed in a certain coordinate system \cite{BR}.

Theories of gravity based on the geometry of distance parallelism
\cite{YJ}$\sim$\cite{Jm} are commonly considered as the closest
alternative to the general relativity theory. Teleparallel gravity
models possess  a number of attractive features both from the
geometrical and physical viewpoints. Teleparallelism is naturally
formulated by gauging external (spacetime) translation and
underlain the Weitzenb$\ddot{o}$ck spacetime characterized by the
metricity condition and by the vanishing of the curvature tensor.
Translations are closely related to the group of general
coordinate transformations which underlies general relativity.
Therefore, the energy-momentum tensor represents the matter source
in the field equations of tetradic theories of gravity like in
general relativity theory.

An important point of teleparallel gravity is that it corresponds
to a gauge theory for the translation group. As a consequence of
translations, any gauge theory including these transformations
will differ from the usual internal gauge models in many ways, the
most significance being the presence of the tetrad field. The
tetrad field can be used to define a linear Weitzenb$\ddot{o}$ck
connection, from which torsion can be defined but no curvature.
Also tetrad field can be used to define a Riemannian metric,  in
terms of which Live-Civita connection is constructed. It is
important to keep in mind that torsion and curvature are
properties of a connection and many different connection can be
defined on the same manifold. The teleparallel equivalent of
general relativity \cite{FW} constitutes an alternative
geometrical description of Einstein's equations.

Tetrad  theories of gravity have been considered long time ago in
connection with attempts to define the energy of gravitational
field \cite{JJTK, M4}. By studying  the properties of the
solutions of Einstein field equations that describe the
gravitational field of an isolated material systems, it is
concluded that a consistent expression for the energy density of
the gravitational field would be given in terms of the quadratic
 form of first-order derivatives of the metric tensor. It
is well known that there exists no covariant, nontrivial
expression constructed out of the metric tensor, both in three and
four dimensions that contain such derivatives. However, covariant
expressions that contain second order derivatives of the tetrad
fields are feasible. Thus it is legitimate to conjecture that the
difficulties regarding the problem of defining the gravitational
energy-momentum is related to the geometrical description of the
gravitational field rather than being an intrinsic drawback of the
theory \cite{SL}.

It is the aim of the present work to find the, asymptotically flat
solutions with spherical symmetry in the tetrad theory of
gravitation for the coupled gravitation and electromagnetic
fields. We obtain three different exact analytic spacetimes in the
tetrad theory of gravitation. To discuss the physical meaning of
these spacetimes we  calculate the energy content of these
solutions using the superpotential method \cite{Mo1,MWHL}.

In \S 2 we derive the field equations of the coupled gravitational
and electromagnetic fields in tetrad theory of gravitation. In \S
3 we first apply the tetrad field given by Robertson \cite{Ro}
with three unknown functions of the radial coordinate $(A(r),B(r)
\ and \ D(r))$ in spherical polar coordinates to the field
equations. We obtain three different exact asymptotically flat
solutions with spherical symmetry in \S 3. In \S 4 the energy
content is calculated using the superpotential method. In \S 5 we
study the asymptotic form  and show that solutions which behave as
${1/\sqrt{r}}$ are not transformed as a four vector as is required
for any consistent energy-momentum complex \cite{Mo66}. The final
section is devoted to the main results and discussion.

Computer algebra system Maple 6 is used in some calculations.

\newsection{The  field equations of the coupled gravitation and electromagnetism  fields}

The dynamics of the gravitational field can be described in the
context of the teleparallel spacetime, where the basic geometrical
entity is the tetrad field ${e^a}_\mu$, (a and $\mu$ are the
$SO(3,1)$ and spacetime indices respectively). Teleparallel
theories of gravity are defined  on the Weitzenb{\rm $\ddot{o}$}ck
spacetime endowed with the affine connection

 \be {\Gamma^\mu}_{\lambda \nu}= {e_k}^\mu \partial_{\nu}
{e^k}_\lambda, \ee where ${\Gamma^\mu}_{\lambda \nu}$ define the
nonsymmetric affine connection coefficients. The curvature tensor
constructed out of Eq. (1), i.e., ${R^\rho}_{\sigma \mu
\nu}(\Gamma)$ vanishes
 identically \cite{HS1}.
  The metric tensor $g_{\mu \nu}$ is given by \be g_{\mu \nu}=
\eta_{k l} {e_\mu}^k {e_\nu}^l \ee with $\eta_{k l}$ is Minkowski
metric $\eta_{k l}=diag(-1,+1,+1,+1)$. Equation (2) leads to
the metricity condition.\\

 The gravitational Lagrangian ${\it L}$ is an invariant constructed from $\gamma_{\mu
\nu \rho}$ and $g^{\mu \nu}$, where $\gamma_{\mu \nu \rho}$ is the
contorsion tensor defined by \be \gamma_{\mu \nu \rho}
\stackrel{\rm def.}{=} e_{k \ \mu} \ {e^k}_{\nu; \ \rho}, \ee
where the semicolon denotes covariant differentiation with respect
to Christoffel symbols. The Lagrangian density which is invariant
under parity operation is given by the form \be {\cal L_G}
\stackrel{\rm def.}{=} (-g)^{1/2} \left( \alpha_1 \Phi^\mu
\Phi_\mu+ \alpha_2 \gamma^{\mu \nu \rho} \gamma_{\mu \nu \rho}+
\alpha_3 \gamma^{\mu \nu \rho} \gamma_{\rho \nu \mu} \right), \ee
where \be g \stackrel{\rm def.}{=} {\rm det}(g_{\mu \nu}), \qquad
and \qquad    \Phi_\mu \stackrel{\rm def.}{=} {\gamma^\rho}_{\mu
\rho}, \ee is the basic vector field.  Here $\alpha_1, \alpha_2,$
and $\alpha_3$ are constants determined by M\o ller such that the
theory coincides with general relativity in the weak fields:

\be \alpha_1=-{1 \over \kappa}, \qquad \alpha_2={\lambda \over
\kappa}, \qquad \alpha_3={1 \over \kappa}(1-\lambda), \ee where
$\kappa$ is the Einstein constant and  $\lambda$ is a free
dimensionless parameter\footnote{Throughout this paper we use the
relativistic units, $c=G=1$ and
 $\kappa=8\pi$.}. The same choice of the parameters was also obtained by Hayashi
 and Nakano \cite{HN}. When this parameter is found to be vanishing Lagrangian (4)
 will reduce to the Lagrangian of teleparallel equivalent of general relativity.

The electromagnetic Lagrangian  density ${\it L_{e.m.}}$ is given
by \be {\it L_{e.m.}}=-\displaystyle{1 \over 4} g^{\mu \rho}
g^{\nu \sigma} F_{\mu \nu} F_{\rho \sigma}, \quad with \quad
F_{\rho \sigma} \quad is \quad given\quad  by\quad
\footnote{Heaviside Lorentz rationalized notations will be used
throughout this paper. We will denote the symmetric part by ( \ ),
for example, $B_{(\mu \nu)}=(1/2)( B_{\mu \nu}+B_{\nu \mu})$ and
the  antisymmetric part by the square bracket [\ ], $B_{[\mu
\nu]}=(1/2)( B_{\mu \nu}-B_{\nu \mu})$ .}  F_{\mu \nu}=
2\partial_{[\mu } A_{\nu]}, \ee and $A_\mu$ is the electromagnetic
potential.

The gravitational and electromagnetic field equations for the
system described by \\ ${\it L_G}+{\it L_{e.m.}}$ are the
following

 \be G_{\mu \nu}(\{\}) +H_{\mu \nu} =
-{\kappa} T_{\mu \nu}, \qquad K_{\mu \nu}=0, \qquad
\partial_\nu \left( \sqrt{-g} F^{\mu \nu} \right)=0, \ee
where $G_{\mu \nu}(\{\})$  is the Einstein tensor and  $H_{\mu
\nu}$ and $K_{\mu \nu}$ are defined by \be H_{\mu \nu}
\stackrel{\rm def.}{=} \lambda \left[ \gamma_{\rho \sigma \mu}
{\gamma^{\rho \sigma}}_\nu+\gamma_{\rho \sigma \mu}
{\gamma_\nu}^{\rho \sigma}+\gamma_{\rho \sigma \nu}
{\gamma_\mu}^{\rho \sigma}+g_{\mu \nu} \left( \gamma_{\rho \sigma
\lambda} \gamma^{\lambda \sigma \rho}-{1 \over 2} \gamma_{\rho
\sigma \lambda} \gamma^{\rho \sigma \lambda} \right) \right],
 \ee
and \be K_{\mu \nu} \stackrel{\rm def.}{=} \lambda \left[
\Phi_{\mu,\nu}-\Phi_{\nu,\mu} -\Phi_\rho \left({\gamma^\rho}_{\mu
\nu}-{\gamma^\rho}_{\nu \mu} \right)+ {{\gamma_{\mu
\nu}}^{\rho}}_{;\rho} \right], \ee and they are symmetric and skew
symmetric tensors, respectively. It can be shown \cite{HS1} that
in spherically symmetric case the antisymmetric part of the field
equations (8), i.e., $K_{\mu \nu}$, implies that the axial-vector
part of the torsion tensor,
$a_\mu=(1/3)\epsilon_{\mu\nu\rho\sigma}\gamma^{\nu\rho\sigma}$,
should be vanishing.  Then the $H_{\mu\nu}$ of (8) vanishes, and
the field equations (8) reduce to the coupled Einstein-Maxwell
equations in teleparallel equivalent of general relativity. The
energy-momentum tensor $T^{\mu \nu}$ is given by  \be T^{\mu
\nu}=g_{\rho \sigma}F^{\mu \rho}F^{\nu \sigma}-\displaystyle{1
\over 4} g^{\mu \nu} g^{\lambda \rho} g^{\epsilon \sigma}
F_{\lambda \epsilon} F_{\rho \sigma} \ee

\newsection{An exact solutions of the coupled
gravitational and electromagnetic field equations with charged
source}

The tetrad space having three unknown functions of radial
coordinate with spherical symmetry   in spherical polar
coordinates, can be written as \cite{Ro}
 \be
\left({e_i}^\mu \right)= \left( \matrix{ A & Dr & 0 & 0
\vspace{3mm} \cr 0 & B \sin\theta \cos\phi & \displaystyle{B \over
r}\cos\theta \cos\phi
 & -\displaystyle{B \sin\phi \over r \sin\theta} \vspace{3mm} \cr
0 & B \sin\theta \sin\phi & \displaystyle{B \over r}\cos\theta
\sin\phi
 & \displaystyle{B \cos\phi \over r \sin\theta} \vspace{3mm} \cr
0 & B \cos\theta & -\displaystyle{B \over r}\sin\theta  & 0 \cr }
\right). \ee  Applying (12) to the field equations (8) one can
obtains a system of non linear differential equations
\cite{Ngr}. Now we are interested in solving these equations:\\
\underline{Special solutions:}\\
i) As a first solution the unknown functions take the following
form \be A(r)=1, \qquad \qquad B(r)=1, \qquad \qquad
D(r)=\displaystyle{ \sqrt{2mr-q^2} \over r^2}.\ee Using (13) in
(12), then line element takes the form \be ds^2=-\left [1-{2m
\over r}+{q^2 \over r^2}\right]dt^2-2\sqrt{{2m \over r}+{q^2 \over
r^2}}drdt+dr^2+r^2d\Omega^2,\ee  with
${d\Omega^2=d\theta^2+\sin^2\theta d\phi^2}$, $m$ and q  are the
mass and the electric charge parameters respectively (both in
length  units). Using the coordinate transformation
\[dT=dt+\displaystyle{D(r) r \over 1-D(r)^2 r^2} dr,\] we can
eliminate the cross term of (14) to obtain \be
ds^2=-\eta(r)dT^2+\displaystyle{dr \over
\eta(r)}+r^2d\Omega^2,\qquad  with \qquad \eta(r)=\left[1-{2m
\over r}+{q^2 \over r^2}\right], \ee which is the  static Reissner
Nordstr$\ddot{o}$m black hole \cite{DGA, NB}. The form of the
vector potential $A_\mu$,  the antisymmetric electromagnetic
 tensor field  $F_{\mu \nu}$ and the energy-momentum tensor are given by
  \be A_t(r)=-\displaystyle{q \over 2\sqrt{\pi} r}, \qquad
  F_{r t}=-\displaystyle{q \over 2\sqrt{\pi} r^2},\qquad
  {T_0}^0={T_1}^1=-{T_2}^2=-{T_3}^3=\displaystyle{q^2 \over 8 \pi r^4}
  \ee
ii) As a second solution the unknown functions take  the following
form \be {\cal A(R)} = \displaystyle{1 \over
\sqrt{1-\displaystyle{2m \over R}+\displaystyle{q^2 \over R^2}}},
\qquad \qquad
 {\cal B(R)} = \sqrt{1-\displaystyle{2m \over
R}+\displaystyle{q^2 \over R^2}} ,  \qquad \qquad {\cal D(R)}=0
\ee where $R=\displaystyle{r \over B}$. Solution (17) gives the
line element (15) which is the  static Reissner Nordstr$\ddot{o}$m
also. Solutions given by (13) and (17) are special solutions.\vspace{0.3cm}\\
iii) As a third solution the unknown functions take the following
form \be {\cal A(R)}= \displaystyle{1 \over \left(1-R {\cal
B}'\right)}, \qquad \qquad
 {\cal D}(R) =\displaystyle{1 \over 1-R {\cal B}'}
\sqrt{\displaystyle{2m \over R^3}+{q^2 \over R^4 }+
\displaystyle{{\cal B}' \over R} \left(R {\cal B}' -2 \right)}.\ee
It is clear from (18) that the third solution depends on the
arbitrary function ${\cal B}$, i.e., we can generate the pervious
solutions (13) and (17) by choosing the arbitrary function ${\cal
B}$ to have the form \be {\cal B(R)} = 1, \qquad and \qquad {\cal
B(R)}={\int{{1 \over R} \left(1-\sqrt{1-\displaystyle{2m \over
R}-\displaystyle{q^2 \over R^2}} \right)}}dR. \ee The line element
of solution (18) takes the form (15) which also is a Reissner
Nordstr$\ddot{o}$m spacetime.  The electromagnetic potential
$A_\mu$, the antisymmetric electromagnetic
 tensor field  $F_{\mu \nu}$  and the
energy-momentum tensor have the form given by (16).

Thus we have three exact solutions of the field equations (8),
each of which leads to the same metric, {\it a static spherically
symmetric  Reissner Nordstr$\ddot{o}$m spacetime in the spherical
polar coordinate and their axial vector part vanishing identically
$a_{\mu}=0$. They practically coincide with the Schwarzschild
solution when the charge $q=0$}.

\newsection{The Energy Associated with each Solution}
In this section we are going to calculate the energy associated
with the above solutions using the superpotential Method. The
superpotential of the M\o ller's theory is given by
\cite{Mo1,MWHL} \be {{\cal U}_\mu}^{\nu \lambda} ={(-g)^{1/2}
\over 2 \kappa} {P_{\chi \rho \sigma}}^{\tau \nu \lambda}
\left[\Phi^\rho g^{\sigma \chi} g_{\mu \tau}
 -\lambda g_{\tau \mu} \gamma^{\chi \rho \sigma}
-(1-2 \lambda) g_{\tau \mu} \gamma^{\sigma \rho \chi}\right], \ee
where ${P_{\chi \rho \sigma}}^{\tau \nu \lambda}$ is \be {P_{\chi
\rho \sigma}}^{\tau \nu \lambda} \stackrel{\rm def.}{=}
{{\delta}_\chi}^\tau {g_{\rho \sigma}}^{\nu \lambda}+
{{\delta}_\rho}^\tau {g_{\sigma \chi}}^{\nu \lambda}-
{{\delta}_\sigma}^\tau {g_{\chi \rho}}^{\nu \lambda} \ee with
${g_{\rho \sigma}}^{\nu \lambda}$ being a tensor defined by \be
{g_{\rho \sigma}}^{\nu \lambda} \stackrel{\rm def.}{=}
{\delta_\rho}^\nu {\delta_\sigma}^\lambda- {\delta_\sigma}^\nu
{\delta_\rho}^\lambda. \ee The energy is expressed by the surface
integral \cite{Mo2} \be E=\lim_{r \rightarrow
\infty}\int_{r=constant} {{\cal U}_0}^{0 \alpha} n_\alpha dS, \ee
where $n_\alpha$ is the unit 3-vector normal to the surface
element ${\it dS}$.

Now we are in a position to calculate the energy associated with
solutions (18) using the superpotential (20). It is
 clear from (23) that, the only components which contributes to the energy is ${{\cal U}_0}^{0
 \alpha}$. Thus substituting from solution (18) into
 (20) we obtain the following non-vanishing value
 \be
{{\cal U}_0}^{0 \alpha}={2X^\alpha \over \kappa R^3}\left(2M- {q^2
\over R}-R^2{\cal B(R)}'\right).
 \ee
 Substituting from (24) into
(23) we get \be E(R)=2M-{q^2 \over R} -R^2{\cal B(R)}'. \ee It is
clear from (25) that if we use the asymptotic form of the
arbitrary function given by (19) we will get two different values
of the energy content! The most satisfactory one is given when the
arbitrary function take the second value of (19) which reproduce
solution (17).

\newsection{ Lorentz Transformation }
M\o ller \cite{Mo66} required that  any satisfactory
energy-momentum complex
must satisfy the following conditions\\
{\it (1) It must be an affine tensor density which satisfies the
conservation law.\\} {\it (2) For an isolated system the
quantities $P_\mu$ are constant in time and transform as the
covariant components of a 4-vector under linear coordinate
transformations.\\ } {\it (3) The superpotential ${{{\cal
U}_\mu}^{\nu \lambda}}=-{{{\cal U}_\mu}^{ \lambda \nu}}$
transforms as a tensor density of rank 3 under the group of the
spacetime transformations.\\}

Now let us examine if these conditions are satisfied by solutions
(13) and (17). Let us start with  solution (17).  The energy
content of this solution is given up to $O(1/R)$ by
$E(R)=m-\displaystyle{q^2+m^2 \over 2R}$ which is a satisfactory
results.

The asymptotic form of the tetrad (12) with solution (13) up to
$O(1/R^2)$ in the Cartesian coordinate is given by \be
\left({e_i}^\mu \right)= \left( \matrix{ 1+\displaystyle{2m(r) r+4
m^2 \over r^2} &-\displaystyle \sqrt{2m \over r}
\left[1-\displaystyle{ q^2 \over 4mr} \right] \displaystyle{x^a
\over r} \vspace{3mm} \cr -\displaystyle \sqrt{{2m \over r}}
\left(1-\displaystyle{q^2-8m^2 \over 4mr} \right)
\displaystyle{x^a \over r} & {\delta_a}^b
   \cr } \right). \ee The energy associated with this tetrad up
   to $O(1/r)$ is given by $E(r)=2m-\displaystyle{q^2
\over r}$ which is different from that given by solution (17)! Now
   we are going to examine if this solution satisfy M\o ller's
   condition or not?

   The non-vanishing components of the superpotential of the tetrad
   (26) up to $O(1/r^3)$ are
   \ba {{\cal U}_0}^{0 \alpha} \A =\A {4
n^\alpha \over \kappa r^2}\left[m-\displaystyle{q^2 \over 2r}
\right], \qquad {{\cal U}_\gamma}^{\beta 0}= \displaystyle{ 1
\over \kappa r^2} \displaystyle{ \sqrt{2 \over m}} \left[\left(
\displaystyle { m^2 \over \sqrt{r}}+\displaystyle { q^2 \over
8\sqrt{r}} \right) {\delta_\gamma}^\beta+\left( \displaystyle {
3m^2 \over \sqrt{r}}-\displaystyle { 5q^2 \over 8\sqrt{r}} \right)
n^\gamma n^\beta \right]=-{{\cal U}_\gamma}^{0 \beta }\nonu
{{\cal U}_\gamma}^{\beta \alpha} \A=\A {m \over \kappa r^2}
 \left(n^\alpha
{\delta_\gamma}^\beta-n^\beta {\delta_\gamma}^\alpha \right).
 \ea
The energy momentum density is defined by \be
{\tau_\mu}^\nu={{{\cal U}_\mu}^{\nu
 \lambda}}_{,\ \lambda},\ee
 using (27) in (28) then, the non-vanishing components of the energy
 momentum density
are expressed by \be
 {\tau_0}^0 = {2 \over \kappa r^3 }\displaystyle{q^2 \over
r},\qquad  {\tau_a}^0  = {4m n^a \over \kappa r^3} \displaystyle
\sqrt{ 2m \over r},\qquad {\tau_a}^b = {m \over \kappa r^3 }
\left[3  n^a n^b- {\delta_a}^b \right].\ee Here we have neglected
higher order terms of $O(1/r)$. The derivatives of (29) gives \[
{{\tau_\mu}^\nu}_{,\ \nu} =0\] which  means that the conservations
law of energy momentum density is satisfied, i.e., condition (1)
of the above conditions is satisfied.

 Now let us examine if condition (2)  is satisfied or not for the tetrad (26)?
  For this purpose we consider  the Lorentz transformation  \be
\bar{x}^0 = \gamma(x^0+vx^1), \qquad \bar{x}^1=\gamma(x^1+vx^0),
\qquad \bar{x}^2=x^2 \qquad \bar{x}^3=x^3 \ee where $\bar{x}$ is
the rest frame of the observer moving with speed $v$ in the
negative direction of the x-axis and $\gamma$ is given by
$\gamma=\displaystyle{1 \over \sqrt{1-v^2}}.$ Here the speed of
light  is taken to be unity. The energy momentum in a volume
element in the new coordinate is given by  \be {{\bar
\tau}_\mu}^\nu d^3{\bar x}=\displaystyle{\partial x^\rho \over
\partial {\bar x}^\mu} \displaystyle{\partial {\bar x}^\nu \over
\partial  x^\sigma}  {\tau_\rho}^\sigma
 \displaystyle{d^3x \over \gamma}. \ee
 Using equations (30) and (31), the  components ${{\bar \tau}_\mu}^0$ take the form
 \be {{\bar \tau}_\mu}^0
d^3{\bar x}=\displaystyle{\partial x^\rho \over
\partial {\bar x}^\mu} \left( {\tau_\rho}^0 +v{\tau_\rho}^1
\right ) d^3x. \ee Integration of (32)  over the three dimensional
hypersurface with  $ \bar \textrm {t}=const.$ gives \be
\int_{{\bar \textrm{ t}} \textrm{=constant}} {{ \bar \tau}_\mu}^0
d^3{\bar x} = \displaystyle{\partial x^\rho \over
\partial {\bar x}^\mu}\left( \int_{\textrm{t=constant}}\left[{\tau_\rho}^0 + v {
\tau_\rho}^1\right] d^3x \right ).\ee Using (29) gives

\be  \int {\tau_\rho}^1 d^3x=\displaystyle{m \over 3}
{\delta_\rho}^1,\ee and finally we obtain
 \be{\bar P}_\mu = \displaystyle{\partial x^\rho \over
\partial {\bar x}^\mu}\left\{P_\rho+\displaystyle {mv \over
3} {\delta_\rho}^1 \right\}, \ee or the four components, \be {\bar
P}_\mu=\gamma \left\{ -\left(E+\displaystyle{mv^2 \over 3}
\right), v\left(E+\displaystyle{m \over 3} \right),0,0
\right\},\ee  where $E=2m$  when $r\rightarrow \infty$. Equation
(36) shows that the four momentum $P_\mu$ is not transformed  as a
4-vector under Lorentz transformation (30), consequently,
condition (2)  of the above conditions is not satisfied!
Therefore, spherically symmetric solutions which behave as
$\displaystyle{1 \over \sqrt{r}}$, is not a physical one.
\newsection{Discussions}

In this paper we have studied the spherically symmetric solutions
in the tetrad  theory of gravitation using the field equations of
the coupled gravitation and electromagnetic fields. The axial
vector part of the torsion tensor, $a^\mu$ of these solutions is
identically vanishing.

Three different exact analytic solutions  are obtained for the
case of spherical symmetry. They give rise to the same Riemannian
metric, i.e.,  a static Reissner Nordstr$\ddot{o}$m spacetime.
 Solutions (13) and (17) represent a black hole which
contain the de Sitter world instead of a singularity.

It was shown by M\o ller \cite{Mo3} that a tetrad description of a
gravitational field equations allows a more satisfactory treatment
of the energy-momentum complex than does general relativity
theory. Therefore, we have applied the superpotential method given
by  M\o ller \cite{Mo1,MWHL} to calculate the energy content of
the general solution (18). It is shown that the energy depend on
the arbitrary function ${\cal B(R)}$. Therefore, the two solutions
(13) and (17) give two different values of energy when the
arbitrary function ${\cal B(R)}$ takes the value (19).

 Now we have three analytical  physical solutions that reproduce the same metric
 spacetime. They, however, yield quite results differ substantially from
 one solution to the other.

Is there any inconsistency between these solutions? To answer this
question we study the conditions given by M\o ller for any
satisfactory energy-momentum complex \cite{Mo66}. Solution (17)
satisfies  all these conditions. However, solution (13) satisfies
only  the first condition, i.e., the conservation law
${{\tau_\mu}^\nu}_{, \ \nu}=0.$ As is clear from (36) that  this
solution  does not satisfy the second  condition, i.e., the four
vector $P_\mu$ is not transformed as a four-vector. Therefore,
solution (13) is not a physical one in spite that its associated
metric gives a well-known metric! Thus we divide the parallel
vector field  into two classes, the one in which the components
$({e^a}_0)$ and $({e^0}_\alpha)$ of the parallel vector fields
$({e^i}_\mu)$ tend to zero faster than $1/\sqrt{r}$ for large r
and the other, in which these components  go to zero as
$1/\sqrt{r}$. We show that
 the first class gives the energy in its standard form while the second class
 gives the value of energy different from the well-known one. This
 is due to the fact that the solutions that belong to the second class
 do not satisfy all the conditions required for any
 satisfactory energy-momentum complex.

\bigskip
\bigskip
\centerline{\Large{\bf Acknowledgements}}

The author would like to thank Professor I.F.I. Mikhail; Ain Shams
University, for his stimulating discussions.

\end{document}